\documentclass[12pt]{article}
\begin{document}

\title{{\bf Finite Canonical Measure for Nonsingular Cosmologies}
\thanks{Alberta-Thy-05-11, arXiv:1103.3699 [hep-th]}}

\author{
Don N. Page
\thanks{Internet address:
profdonpage@gmail.com}
\\
Theoretical Physics Institute\\ 
Department of Physics, University of Alberta\\
Room 238 CEB, 11322 -- 89 Avenue\\
Edmonton, Alberta, Canada T6G 2G7
}

\date{2011 June 1}

\maketitle
\large
\begin{abstract}
\baselineskip 20 pt

The total canonical (Liouville-Henneaux-Gibbons-Hawking-Stewart) measure
is finite for completely nonsingular
Friedmann-Lema\^{\i}tre-Robertson-Walker classical universes with a
minimally coupled massive scalar field and a positive cosmological
constant.  For a cosmological constant very small in units of the square
of the scalar field mass, most of the measure is for nearly de Sitter
solutions with no inflation at a much more rapid rate.  However, if one
restricts to solutions in which the scalar field energy density is ever
more than twice the equivalent energy density of the cosmological
constant, then the number of e-folds of rapid inflation must be large,
and the fraction of the measure is low in which the spatial curvature is
comparable to the cosmological constant at the time when it is
comparable to the energy density of the scalar field. 

The measure for such classical FLRW$\Lambda$-$\phi$ models with both a
big bang and a big crunch is also finite.  Only the solutions with a big
bang that expand forever, or the time-reversed ones that contract from
infinity to a big crunch, have infinite measure.

\end{abstract}

\normalsize

\baselineskip 14 pt

\newpage

\section{Introduction}

Starting with the Louiville measure, the procedure of Henneaux
\cite{Hen} and, in more detail, of Gibbons, Hawking, and Stewart
\cite{GHS}, provides a natural canonical measure on the set of classical
universes.  For a minisuperspace model of a $k=+1$
Friedmann-Lema\^{\i}tre-Robertson-Walker (FLRW) geometry (homogeneous
and isotropic three-sphere spatial sections) with a time-dependent scale
size $a(t)$ and a single minimally coupled homogeneous massive scalar
field $\phi(t)$, the total measure is infinite, and Hawking and I showed
\cite{HP} that all but a finite measure have arbitrarily small spatial
curvature $1/a^2$ (or arbitrarily large size $a$) at any fixed positive
value of the energy density of the scalar field.  However, we also
showed \cite{HP} that both the set of inflationary solutions and the set
of noninflationary solutions have infinite measure, with an ambiguous
ratio:  the Louiville-Henneaux-Gibbons-Hawking-Stewart canonical
classical measure gives an ambiguous prediction for the probability of
inflation.

Gibbons and Turok \cite{GT} sought to remove this ambiguity by
identifying universes in which the spatial curvature $1/a^2$ is too
small to be distinguished, though in another way of looking at it, it is
not clear why one should be justified in identifying universes that have
scale sizes $a$ (and hence also spatial volumes that are proportional to
$a^3$) that are so large and different.  Indeed, Turok has abandoned
this approach and is working on another \cite{Tur}.  In both of these
procedures, one gets a classical probability of inflation that goes
approximately inversely with the volume expansion factor during
inflation, i.e., roughly as $\exp{(-3N)}$, where $N$ is the number of
e-folds of inflation \cite{GT,Tur}.

Here a very simple alternative restriction of the set of classical
solutions is considered that does not depend on any identifications of
what might not be observationally distinguished and does not depend on
any arbitrary choice of finite ranges.  In particular, I consider the
set of classical FLRW solutions that are completely nonsingular, with
neither a big bang nor a big crunch.  For FLRW solutions with a
homogeneous minimally coupled massive scalar field and with a
cosmological constant, nonsingular solutions with positive canonical
measure occur only for the $k=+1$ FLRW geometries (allowing $a(t)$ to
have a minimum value) and for a positive cosmological constant $\Lambda$
(allowing the universe to expand forever in both directions of time). 
(There is an uncountable set of perpetually bouncing solutions for a
homogeneous massive scalar field minimally coupled to a $k=+1$ FLRW
geometry with $\Lambda = 0$ \cite{dnpcqg}, but this apparently fractal
set is discrete and so has zero canonical measure.)  This restriction of
positive measure for nonsingular solutions to $k=+1$ and $\Lambda > 0$
is also true for any homogeneous minimally coupled scalar field with a
canonical kinetic term and a potential term with one single extremum
that is a minimum of zero value, but here we shall focus on the
homogeneous free massive inflaton scalar field $\phi(t)$ of mass $m$,
taking the cosmological constant to be $\Lambda \equiv 3/b^2 \equiv
3m^2\lambda > 0$ with a characteristic length scale $b$ and a
dimensionless rescaled cosmological constant $\lambda$.

This paper shows that the set of such nonsingular classical FLRW
universes has finite canonical measure.  Therefore, if one restricts to
such cosmologies with neither a big bang nor a big crunch, one can get
unambiguous finite fractions for any subsets one chooses.

\baselineskip 15 pt

\section{Friedmann-Lema\^{\i}tre-Robertson-Walker closed model with
massive scalar field and cosmological constant}

The Friedmann-Lema\^{\i}tre-Robertson-Walker (FLRW) closed model with
massive scalar field and possibly also with a cosmological constant has
been analyzed many times previously
\cite{PF,Star,Hawk,dnpcqg,BGKZ,BK,KKT,KKT2,CS,KKST,Symbounce},
so many of the dynamical equations I shall give below have been
previously given, along with much of the qualitative behavior.

The $k=+1$ FLRW spacetime metric is
\begin{equation}
ds^2 = -N^2(t)dt^2 + A^2(t)d\Omega_3^2
     = \frac{1}{m^2}(-n^2(t)dt^2 + a^2(t)d\Omega_3^2),
\label{eq:a}
\end{equation}
where $N(t)$ is the lapse function, $A(t)$ is the physical scale size
(which is what is usually called $a(t)$, but I am reserving that for the
rescaled scale size), $d\Omega_3^2$ is the metric on a unit 3-sphere
that has volume $2\pi^2$, $n(t) \equiv mN(t)$ is a rescaled lapse
function that is dimensionless if $t$ is taken to be dimensionless, and
$a(t) \equiv mA(t)$ is a rescaled scale size that is also dimensionless.

Using units in which $\hbar = c = 1$, but writing Newton's constant $G
\equiv m_{\mathrm{Pl}}^{-2}$ or the Planck mass $m_{\mathrm{Pl}} \equiv
G^{-1/2} \equiv \sqrt{\hbar c/G}$ explicitly, the Lorentzian action is
(cf.\ \cite{dnpcqg,Symbounce}, but note that here I am using $A(t)$ for
the physical scale factor and $a(t)$ for the dimensionless rescaled
scale factor, unlike the $a(t)$ and $r(t)$ used in \cite{Symbounce} for
those two respective quantities)
\begin{eqnarray}
S &=& \int Ndt 2\pi^2 A^3 \left\{\frac{3}{8\pi G}
        \left[-\left(\frac{1}{NA}\frac{dA}{dt}\right)^2
	      +\frac{1}{A^2}-\frac{\Lambda}{3}\right]
	+\frac{1}{2}\left(\frac{1}{N}\frac{d\phi}{dt}\right)^2
	-\frac{1}{2}m^2\phi^2\right\} \nonumber \\
  &=& \frac{3\pi}{4G}\int Ndt A^3 \left\{
        -\left(\frac{1}{NA}\frac{dA}{dt}\right)^2
	+\left(\frac{1}{N}\frac{d\varphi}{dt}\right)^2
	+\frac{1}{A^2}-\frac{1}{b^2}-m^2\varphi^2\right\} \nonumber \\
  &=& \frac{1}{2}\left(\frac{3\pi}{2Gm^2}\right)\int ndt a^3 \left\{
        -\left(\frac{1}{na}\frac{da}{dt}\right)^2
	+\left(\frac{1}{n}\frac{d\varphi}{dt}\right)^2
	+\frac{1}{a^2}-\lambda-\varphi^2\right\} \nonumber \\
  &=& \frac{1}{2}S_0\int ndt e^{3\alpha}
       \left[-n^{-2}(\dot{\alpha}^2-\dot{\varphi}^2)
        +e^{-2\alpha}-\lambda-\varphi^2\right] \nonumber \\
  &=& \frac{1}{4}S_0\int dt \left\{-\bar{n}^{-1}\dot{u}\dot{v}
     +\bar{n}\left[1-\left(\lambda+\frac{1}{16}\ln^2{\frac{v}{u}}\right)
       \sqrt{uv} \right] \right\} \nonumber \\
  &=& \frac{1}{2}S_0\int dt \left\{-\frac{4}{9}n^{-1}\dot{U}\dot{V}
        + n (UV)^{1/3} 
          \left[1-\left(\lambda+\frac{1}{9}\ln^2{\frac{V}{U}}\right)
             (UV)^{2/3} \right] \right\} \nonumber \\	
  &=& \frac{1}{2}\int ndt 
       \left[\left(\frac{1}{n}\frac{ds}{dt}\right)^2-\hat{V}\right]
      = \frac{1}{2}\int dt 
         \left[\frac{1}{\hat{n}}\left(\frac{d\hat{s}}{dt}\right)^2
                          -\hat{n}\right],
\label{eq:b}
\end{eqnarray}
where $b\equiv \sqrt{3/\Lambda}$ is the radius of the throat of pure de
Sitter with the same value of the cosmological constant, $\lambda \equiv
\Lambda/(3m^2) \equiv 1/(mb)^2$ is a dimensionless measure of the
cosmological constant in units given by the mass of the inflaton, $a
\equiv e^{\alpha} \equiv mA \equiv (uv)^{1/4} \equiv (UV)^{1/3}$ and
$\varphi \equiv \sqrt{4\pi G/3}\phi \equiv (1/4)\ln{(v/u)} \equiv
(1/3)\ln{(V/U)}$ are dimensionless forms of the scale factor and
inflaton scalar field, $u = e^{2\alpha-2\varphi} = a^2 e^{-2\varphi}$
and $v = e^{2\alpha+2\varphi} = a^2 e^{+2\varphi}$ are a convenient
choice of null coordinates on the minisuperspace (see, e.g.,
\cite{conf}), $U=u^{3/4}=e^{(3/2)\alpha-(3/2)\varphi} = a^{3/2}
e^{-2\varphi}$ and $V=v^{3/4}=e^{(3/2)\alpha+(3/2)\varphi} = a^{3/2}
e^{+2\varphi}$ are an alternative choice of null coordinates, an overdot
represents a derivative with respect to $t$, $S_0 \equiv (3\pi)/(2Gm^2)
= (3\pi/2)(m_{\mathrm{Pl}}/m)^2$, the DeWitt metric \cite{DW} on the
minisuperspace is
\begin{equation}
ds^2 = S_0 e^{3\alpha}(-d\alpha^2 + d\varphi^2)    
     = -\frac{1}{4}S_0(uv)^{-1/4}dudv = -\frac{4}{9}S_0dUdV,
\label{eq:c}
\end{equation}
the `potential' on the minisuperspace is
\begin{equation}
\hat{V} = S_0 e^{3\alpha}(\varphi^2+\lambda-e^{-2\alpha})
   = -S_0(uv)^{1/4}\left[1-\left(\lambda+\frac{1}{16}\ln^2{\frac{v}{u}}
       \right)\sqrt{uv} \right],
\label{eq:d}
\end{equation}
alternative rescaled lapse functions are $\bar{n} = 2(uv)^{1/4}n = 2an =
2m^2AN$ and $\hat{n} \equiv n\hat{V} = mN\hat{V}$, and the conformal
minisuperspace metric is
\begin{eqnarray}
d\hat{s}^2 &=& \hat{V}ds^2 = S_0^2 e^{6\alpha}
       (\varphi^2+\lambda-e^{-2\alpha})(-d\alpha^2 + d\varphi^2)
                    \nonumber \\
           &=& \frac{1}{4} S_0^2
            \left[1-\left(\lambda+\frac{1}{16}\ln^2{\frac{v}{u}}\right)
             \sqrt{uv} \right] du dv \nonumber \\
	   &=& \frac{4}{9} S_0^2 (UV)^{1/3}
	    \left[1-\left(\lambda+\frac{1}{9}\ln^2{\frac{V}{U}}\right)
             (UV)^{2/3} \right].
\label{eq:e}
\end{eqnarray}
The null coordinates $u$ and $v$ are chosen so that as one approaches
the null boundaries of the minisuperspace, at $u \geq 0$, $v=0$ where
$a=0$ with $\varphi = -\infty$ (except at $u=v=0$, where $\varphi$ can
have any value), and at $u=0$, $v \geq 0$ where $a=0$ with $\varphi =
+\infty$ (again except at $u=v=0$), the conformal minisuperspace metric
Eq.\ (\ref{eq:e}) approaches $(1/4)S_0^2dudv$, so that $u$ and $v$ are
$2/S_0$ times null coordinates that are the local analogues of
orthonormal Minkowski coordinates near the boundaries.

The Hamiltonian constraint equation and independent equation of motion
can now be written as
\begin{eqnarray}
 &&\left(\frac{1}{Na}\frac{dA}{dt}\right)^2
=\left(\frac{1}{N}\frac{d\varphi}{dt}\right)^2
 +m^2\varphi^2 + \frac{1}{b^2} - \frac{1}{A^2}, \nonumber \\
 &&\frac{1}{N}\frac{d}{dt}\left(\frac{1}{N}\frac{d\varphi}{dt}\right)
+\left(\frac{3}{NA}\frac{da}{dt}\right)
       \left(\frac{1}{N}\frac{d\varphi}{dt}\right) + m^2\varphi^2 = 0,
\label{eq:f}
\end{eqnarray}
for general lapse function from the second form of the action above,
\begin{eqnarray}
&&\dot{a}^2 = a^2(\dot{\varphi}^2 + \varphi^2 + \lambda) - 1,
   \nonumber \\
&&\ddot{\varphi} + 3\frac{\dot{a}}{a}\dot{\varphi} + \varphi = 0,
\label{eq:g}
\end{eqnarray}
from the third form of the action with $n=1$, which will henceforth be
assumed unless otherwise indicated (e.g., by including the lapse $N$ or
$n$ explicitly in a formula),
\begin{eqnarray}
&&\dot{\alpha}^2 - \dot{\varphi}^2 = \varphi^2+\lambda-e^{-2\alpha},
   \nonumber \\
&&\ddot{\varphi} + 3\dot{\alpha}\dot{\varphi} + \varphi = 0,
\label{eq:h}
\end{eqnarray}
for the fourth form of the action, and
\begin{eqnarray}
&&\dot{u}\dot{v} = -4\sqrt{uv}
 \left[1-\left(\lambda+\frac{1}{16}\ln^2{\frac{v}{u}}\right)
   \sqrt{uv} \right],
   \nonumber \\
&&\frac{\ddot{U}}{U}-\frac{\ddot{V}}{V} = \ln{\frac{V}{U}},
\label{eq:hh}
\end{eqnarray}
for the minisuperspace null coordinates $u=e^{2\alpha-2\varphi}$ and
$v=e^{2\alpha+2\varphi}$, and the alternative null coordinates
$U=u^{3/4}=e^{(3/2)\alpha-(3/2)\varphi}$ and
$V=v^{3/4}=e^{(3/2)\alpha+(3/2)\varphi}$.  This is with unit value for
the rescaled lapse function, $n=1$, but the last two equations appear
simpler directly in terms of $u$ and $v$ if for just these equations we
use the rescaled lapse $\bar{n} = 1$ or $n = 1/(2a) = (1/2)e^{-\alpha} =
(1/2)(uv)^{-1/4} = (1/2)(UV)^{-1/3}$, which gives
\begin{eqnarray}
&&\dot{u}\dot{v} = 
-1+\left(\lambda+\frac{1}{16}\ln^2{\frac{v}{u}}\right)\sqrt{uv},
   \nonumber \\
&&\frac{\ddot{u}}{u}-\frac{\ddot{v}}{v}
   = \frac{1}{4\sqrt{uv}}\ln{\frac{v}{u}},
\label{eq:hhh}
\end{eqnarray}

Although they are redundant equations, one may readily derive from Eqs.\
(\ref{eq:g}) and (\ref{eq:h}) that
\begin{equation}
\ddot{a} = a(\varphi^2-2\dot{\varphi}^2+\lambda)
         = a\left(\frac{\dot{a}^2+1}{a^2}-2\dot{\varphi}^2\right)
\label{eq:g2}
\end{equation}
and
\begin{equation}
\ddot{\alpha} = e^{-2\alpha}-3\dot{\varphi}^2 
\label{eq:h2}
\end{equation}
when $n=1$.  Then when neither side of the constraint (first) equation
part of Eqs. (\ref{eq:h}) vanishes (e.g., when $\hat{V}\neq 0$), and
when $\dot{\varphi}\neq 0$, one may define $f' \equiv df/d\varphi =
\dot{f}/\dot{\varphi}$ and reduce Eqs.\ (\ref{eq:h}) to the single
second-order differential equation (cf.\ \cite{dnpcqg})
\begin{equation}
\alpha'' = \frac{(\alpha'^2-1)
            (\varphi \alpha' + 3\varphi^2 + 3\lambda - 2e^{-2\alpha})}
	    {\varphi^2+\lambda-e^{-2\alpha}}.
\label{eq:i}
\end{equation}
Alternatively, when $\hat{V}\neq 0$ (or equivalently $\dot{\alpha}^2
\neq \dot{\varphi}^2$), but when $\dot{\alpha} \neq 0$ instead of
$\dot{\varphi}\neq 0$, one can write
\begin{equation}
\frac{d^2\varphi}{d\alpha^2} = 
\frac{(d\varphi/d\alpha)^2-1}{\varphi^2+\lambda-e^{-2\alpha}}
 \left[\left(3\varphi^2 + 3\lambda - 2e^{-2\alpha}\right)
 \frac{d\varphi}{d\alpha} + \varphi\right].
\label{eq:j}
\end{equation}

Yet another way to get the equations of motion is to note that the
seventh (penultimate) form of the action from Eq.\ (\ref{eq:b}) gives
the trajectories of a particle of mass-squared $\hat{V}$ in the DeWitt
minisuperspace metric $ds^2$, and the eighth and final form of the
action gives timelike geodesics in the conformal minisuperspace metric
$d\hat{s}^2 = \hat{V}ds^2$.  When one goes to the gauge $\hat{n} = 1$,
then $(d\hat{s}/dt)^2 = -1$, so that along the classical timelike
geodesics of $d\hat{s}^2$, the Lorentzian action is $S = -\int dt =
-\int\sqrt{-d\hat{s}^2}$, minus the proper time along the timelike
geodesic of $d\hat{s}^2$.  However, one must note that the conformal
metric $d\hat{s}^2 = \hat{V}ds^2$ is singular at $\hat{V}=0$, that is at
$\varphi^2 + \lambda = e^{-2\alpha} \equiv 1/a^2 \equiv 1/(mA)^2$,
whereas there is no singularity in the DeWitt metric $ds^2$ or the
spacetime metric along this hypersurface (curve) in the two-dimensional
minisuperspace $(\alpha,\varphi)$ under consideration.  The second-order
differential equations (\ref{eq:i}) and (\ref{eq:j}) also break down at
$\hat{V}=0$ and must be supplemented by the continuity of $\dot{\alpha}$
and of $\dot{\varphi}$ (in a gauge in which $n \neq 0$ is continuous
there) across the $\hat{V}=0$ hypersurface (curve).

To get reasonable numbers for the dimensionless constants in these
equations, I shall follow \cite{Symbounce} and set $m \approx 1.5\times
10^{-6} G^{-1/2} \approx 7.5\times 10^{-6} (8\pi G)^{-1/2}$
\cite{Lindebook,LL}, so the prefactor of the action becomes $S_0 \equiv
(3\pi)/(2Gm^2) = (3\pi/2)(m_{\mathrm{Pl}}/m)^2 \approx 2\times
10^{12}$, and the dimensionless measure of the cosmological constant is
$\lambda \equiv \Lambda/(3m^2) \equiv 1/(mb)^2 \approx 5\times
10^{-111}$.  Thus $\lambda$ may be taken to be extremely tiny.

The constrained Hamiltonian system for this $k=+1$ FLRW$\Lambda$-$\phi$
model universe has an unconstrained $2d$-dimensional phase space
$\Gamma_d$ with $d=2$ that may be
labeled by the coordinates $Q^i$ and conjugate momenta $P_i$, which may
be chosen to be any of the following sets:
\begin{equation}
\{A,\ \phi,\ p_A = -\frac{3\pi}{2Gm}\frac{A}{N}\frac{dA}{dt},\ 
p_\phi = +\frac{2\pi^2 A^3}{N}\frac{d\phi}{dt}\},
\label{eq:coor1}
\end{equation}
\begin{equation}
\{a,\ \varphi,\ p_a = -S_0\frac{a}{n}\frac{da}{dt},\ 
p_\varphi = +S_0\frac{a^3}{n}\frac{d\varphi}{dt}\},
\label{eq:coor2}
\end{equation}
\begin{equation}
\{\alpha,\ \varphi,\ 
p_\alpha = -S_0\frac{e^{3\alpha}}{n}\dot{\alpha},\ 
p_\varphi = +S_0\frac{e^{3\alpha}}{n}\dot{\varphi}\},
\label{eq:coor3}
\end{equation}
\begin{equation}
\{u,\ v,\ 
p_u = -\frac{S_0}{4\bar{n}}\dot{v},\ 
p_v = -\frac{S_0}{4\bar{n}}\dot{u}\},
\label{eq:coor4}
\end{equation}
\begin{equation}
\{U,\ V,\ 
p_U = -\frac{2S_0}{9n}\dot{V},\ 
p_V = -\frac{2S_0}{9n}\dot{U}\}.
\label{eq:coor5}
\end{equation}

The Hamiltonian on this phase space is then
\begin{eqnarray}
H &=& -\frac{GN}{3\pi A}p_A^2 + \frac{N}{4\pi^2 A^3}p_\phi^2
      -\frac{3\pi N}{4G}A +\frac{\pi\Lambda N}{4G}A^3 
      +\pi^2 m^2 N A^3 \phi^2  \nonumber \\
  &=& \frac{Gm^2n}{3\pi}\left(-\frac{p_a^2}{a}
                              +\frac{p_\varphi^2}{a^3}\right)
      +\frac{3\pi n}{4G m^2}\left(-a+\lambda a^3+a^3\varphi^2\right)
           \nonumber \\
  &=& \frac{1}{2S_0} n e^{-3\alpha}
       (-p_\alpha^2 + p_\varphi^2)
      +\frac{S_0}{2} n e^{3\alpha}
        (-e^{-2\alpha} + \lambda + \varphi^2)
	   \nonumber \\
  &=& -\frac{4}{S_0}\bar{n} p_u p_v
      -\frac{S_0}{4}\bar{n} 
        \left[1-\left(\lambda+\frac{1}{16}\ln^2{\frac{v}{u}}\right)
         \sqrt{uv} \right] \nonumber \\
  &=& -\frac{9}{2S_0} n p_U p_V
      - \frac{1}{2}S_0 n (UV)^{1/3} 
         \left[1-\left(\lambda+\frac{1}{9}\ln^2{\frac{V}{U}}\right)
          (UV)^{2/3} \right] \nonumber \\	   	   
  &=& \frac{1}{2}S_0\left[\frac{-a\dot{a}^2 + a^3\dot{\varphi}^2}{n}
                       +n\left(-a+\lambda a^3+a^3\varphi^2\right)\right]
		          \nonumber \\
  &=& \frac{1}{2}S_0 n e^{3\alpha}
       \left[-\left(\frac{\dot{\alpha}}{n}\right)^2
             +\left(\frac{\dot{\varphi}}{n}\right)^2
	     -e^{-2\alpha} + \lambda + \varphi^2 \right] \nonumber \\
  &=& \frac{1}{4} S_0 \left\{-\bar{n}^{-1}\dot{u}\dot{v} -\bar{n}
          \left[1-\left(\lambda+\frac{1}{16}\ln^2{\frac{v}{u}}\right)
           \sqrt{uv} \right] \right\} \nonumber \\
  &=& \frac{1}{2} S_0 \left\{-\frac{4}{9}n^{-1}\dot{U}\dot{V}
         -n(UV)^{1/3}
          \left[1-\left(\lambda+\frac{1}{9}\ln^2{\frac{V}{U}}\right)
          (UV)^{2/3} \right] \right\}.
	\label{eq:Hamiltonian}
\end{eqnarray}

The last four expressions are not in the canonical form as functions of
the coordinates and momenta but are given to express the value of the
Hamiltonian in terms of the coordinates and their time derivatives. 
Because the Hamiltonian constraint equation, obtained by varying the
action $S$ of Eq.\ ({\ref{eq:b}}) with respect to the lapse function $n$
or $\bar{n}$, is $H=0$, these last four expressions for $H$ can be
easily seen to lead to the first equations in each of Eqs.\
(\ref{eq:g}), (\ref{eq:h}), (\ref{eq:hh}), and (\ref{eq:hhh}) when one
chooses the rescaled lapse function $n$ to be 1 for Eqs.\ (\ref{eq:g}),
(\ref{eq:h}), and (\ref{eq:hh}) and chooses $\bar{n}=1$ in Eq.\
(\ref{eq:hhh}).  One can also write the Hamiltonian constraint equation
$H=0$ directly n terms of the canonical coordinates and momenta as
\begin{equation}
a^2 p_a^2 - p_\varphi^2 =
S_0^2 a^6 \left(\lambda + \varphi^2 - a^{-2} \right),
\label{eq:mom1}
\end{equation}
\begin{equation}
p_\alpha^2 - p_\varphi^2 =
S_0^2 e^{6\alpha} \left(\lambda + \varphi^2 - e^{-2\alpha} \right),
\label{eq:mom2}
\end{equation}
\begin{equation}
p_u p_v =
-\frac{1}{16} S_0^2 
  \left[1-\left(\lambda+\frac{1}{16}\ln^2{\frac{v}{u}}\right)
   \sqrt{uv} \right],
\label{eq:mom3}
\end{equation}
\begin{equation}
p_U p_V =
-\frac{1}{9} S_0^2 (UV)^{1/3}
          \left[1-\left(\lambda+\frac{1}{9}\ln^2{\frac{V}{U}}\right)
           (UV)^{2/3} \right].
\label{eq:mom4}
\end{equation}

\section{The canonical measure for $k=+1$ FLRW$\Lambda$-$\phi$}

The natural symplectic structure for the $k=+1$ FLRW$\Lambda$-$\phi$
constrained Hamiltonian system is the closed nondegenerate differential
2-form
\begin{eqnarray}
\omega_n &=&\omega_2 = dP_i \wedge dQ^i \nonumber \\
&=& dp_A \wedge dA + dp_\phi \wedge d\phi
 = -S_0 m^2 Ad\dot{A}\wedge dA + 2\pi^2 m d(A^3\dot{\phi})\wedge d\phi
               \nonumber \\
&=& dp_a \wedge da + dp_\varphi \wedge d\varphi
 = S_0 \left(-a d\dot{a} \wedge da
   + d(a^3\dot{\varphi}) \wedge d\varphi \right)
    \nonumber \\
&=& dp_\alpha \wedge d\alpha + dp_\varphi \wedge d\varphi
 = S_0 e^{3\alpha}
 (-d\dot{\alpha} \wedge d\alpha+d\dot{\varphi} \wedge d\varphi
   +3\dot{\varphi} d\alpha \wedge d\varphi) \nonumber \\
&=& dp_u \wedge du + dp_v \wedge dv
 = -\frac{S_0}{8}(uv)^{1/4}\left[d\dot{v} \wedge du + d\dot{u} \wedge dv
 +\frac{1}{4}\left(\frac{\dot{v}}{v}-\frac{\dot{u}}{u}\right)du\wedge dv
 \right] \nonumber \\
&=& dp_U \wedge dU + dp_V \wedge dV
 = -\frac{2}{9}S_0\left(d\dot{V}\wedge dU + d\dot{U}\wedge dV \right),
\label{eq:symform}
\end{eqnarray}
where for the expressions in terms of the time derivatives, I have used
the default option $n=1$.

When this is pulled back to the $H=0$ constraint hypersurface of
dimension $2d-1 = 3$ in the unconstrained phase space $\Gamma_d =
\Gamma_2$ of dimension $2d = 4$ and further pulled back to an
initial-data surface $\Gamma_{d-1} = \Gamma_1$ of dimension $2d-2 = 2$
that is transverse to the Hamiltonian flow in the 3-dimensional
constraint hypersurface, it gives the symplectic structure differential
form $\omega \equiv \omega_{d-1} = \omega_1$ on that 2-dimensional
initial data surface.  Since $d-1 = 1$, it is the first power of this
symplectic structure form that gives the canonical
Liouville-Henneaux-Gibbons-Hawking-Stewart measure or volume (area)
element $\Omega_{n-1} = \Omega_1 = \omega$ on the initial data surface
$\Gamma_1$ \cite{GHS}.  That is, if a bunch of orbits $B$ intersects an
initial-data surface $\Sigma$ in the region $S$, the canonical measure
of that bunch of orbits is $\mu(B) = \int_S \omega$.  As Gibbons,
Hawking, and Stewart show \cite{GHS}, this measure is preserved as one
follows the bunch $B$ to where it intersects a different initial-data
surface $\Sigma'$ in the region $S'$, giving the same measure $\mu(B) =
\int_{S'} \omega$.

Here we are restricting to nonsingular cosmologies,
Friedmann-Lema\^{\i}tre-Robertson-Walker universes that have neither a
big bang nor a big crunch, so the scale factor $A$ or $a$ never goes to
zero.  Except for a discrete set of zero canonical measure
\cite{dnpcqg}, all of these solutions will contract from infinite $a$ at
infinite past time and re-expand to infinite $a$ at infinite future
time.  Therefore, they will each have a global minimum for $a =
e^\alpha$, which I shall label $a_m \equiv \exp{(\alpha_m)}$, where
$da/dt \equiv \dot{a} = e^\alpha\dot{\alpha} = 0$ and hence $p_A = p_a =
p_\alpha = 0$.  Let $\varphi_m$, $\dot{\varphi}_m$, $\ddot{\alpha}_m$,
and $p_{\varphi_m}$ be the values of $\varphi$, $\dot{\varphi}$,
$\ddot{\alpha}$, and $p_\varphi$ at this global minimum for $a$ and
hence also for $\alpha$. 

The Hamiltonian constraint $H=0$, given by the first Eq.\ (\ref{eq:g})
with both sides equal to zero, implies that (using the default setting
of the rescaled lapse function as $n=1$)
\begin{equation}
a_m = (\varphi_m^2 + \dot{\varphi}_m^2 + \lambda)^{-1/2},
\label{eq:k1}
\end{equation}
or
\begin{equation}
\alpha_m = 
-\frac{1}{2}\ln{(\varphi_m^2 + \dot{\varphi}_m^2 + \lambda)},
\label{eq:k2}
\end{equation}
Therefore, initial data are given by values of $\varphi_m$ and
$\dot{\varphi}_m$, with $a_m = a_m(\varphi_m,\dot{\varphi}_m)$ given by
Eq.\ (\ref{eq:k1}), and then one has $\dot{a} = 0$ automatically at this
point in the constrained phase space.

One can further readily see (cf.\ \cite{KKT,KKT2})that at an extremum
for $a$, where $\dot{a} = 0$, that one has
\begin{equation}
\frac{\ddot{a}}{a}=\ddot{\alpha}= 3\varphi^2 + 3\lambda - 2e^{-2\alpha}
= \varphi^2 - 2\dot{\varphi}^2  + \lambda.
\label{eq:k3}
\end{equation}
Therefore, for an extremum to be at least a local minimum, one needs
$a_m \leq (2/3)^{1/2}(\varphi_m^2 + \lambda)^{-1/2}$ or
$\dot{\varphi}_m^2 \leq (\varphi_m^2 + \lambda)/2$, though there are
additional conditions for such a local minimum to be a global minimum.

Since Eq.\ (\ref{eq:k1}) gives a unique value for $a_m$ for each set of
real values for $\varphi_m$ and $\dot{\varphi}_m$, it na\"{\i}vely
appears that there are no constraints on $\varphi_m$ and
$\dot{\varphi}_m$.  However, different sets of $\varphi_m$ and
$\dot{\varphi}_m$ can lead to the same solution, since a solution may
have more than one point along its trajectory (more than one time) where
$\dot{a} = 0$, only one of which may be a global minimum in the generic
case in which there are not more than one time with the same global
minimum value of $a(t)$.  Therefore, counting all possibilities for
$\varphi_m$ and $\dot{\varphi}_m$ overcounts the trajectories that have
$\dot{a} = 0$ somewhere along them.  Furthermore, singular trajectories,
with $a$ going to zero in the past or future, may also have points where
$\dot{a} = 0$ and thus be counted if one counts all possible pairs of
$\varphi_m$ and $\dot{\varphi}_m$.  In the next Section we shall look at
the restrictions on $\varphi_m$ and $\dot{\varphi}_m$ in order that this
pair correspond to a global minimum of $a$ rather than some other
extremum like a local minimum that is not a global minimum, or either a
local or a global maximum.  However, first we shall show that the total
canonical measure of all solutions with any local nonzero extremum for
$a$ (at $a=a_m$ or $\alpha = \alpha_m$ where $\dot{a}=0$ and
$\dot{\alpha} = 0$) has finite measure.

It is convenient to define two new variables $\beta$ and $\theta$ so
that (with rescaled lapse function $n=1$ as usual)
\begin{eqnarray}
\varphi &=& e^{-\beta} \cos{\theta}, \nonumber \\
\dot{\varphi} &=& e^{-\beta} \sin{\theta}.
\label{eq:bt}
\end{eqnarray}
The Hamiltonian constraint equation Eq.\ (\ref{eq:h}) then becomes
\begin{equation}
\dot{\alpha}^2 = \lambda + e^{-2\beta} - e^{-2\alpha}.
\label{eq:bc}
\end{equation}
One can also easily calculate that the time derivatives of $\beta$ and
$\theta$ are
 \begin{equation}
\dot{\beta} = 3\dot{\alpha}\sin^2{\theta} 
            = \frac{3}{2}\dot{\alpha}(1-\cos{2\theta}),
\label{eq:betadot}
\end{equation}
\begin{equation}
\dot{\theta} = -1 - 3\dot{\alpha}\sin{\theta}\cos{\theta}
             = -1 - \frac{3}{2}\dot{\alpha}\sin{2\theta}.
\label{eq:thetadot}
\end{equation}

When averaged over one period of $\theta$ in a regime in which the
scalar field oscillates rapidly relative to the expansion (so that the
time-average of the scalar field stress-energy tensor is approximately
that of pressureless dust), $\beta$ changes by approximately $3/2$ as
much as $\alpha$, so it is convenient to define a total rationalized
dimensionless `mass' (twice the energy density multiplied by the volume
and divided by $S_0$ and by the scalar field mass $m$) that is nearly
constant in the dustlike regime \cite{Symbounce}:
\begin{equation}
M \equiv e^{3\alpha-2\beta} \equiv a^3(\varphi^2 + \dot{\varphi}^2),
\label{eq:M}
\end{equation}
obeying
\begin{equation}
\dot{M} = 3M\dot{\alpha}\cos{2\theta}
\label{eq:Mdot}
\end{equation}
or
\begin{equation}
\frac{d\ln{M}}{d\alpha} = 3\cos{2\theta}.
\label{eq:dMdalpha}
\end{equation}

The symplectic structure 2-form $\omega_n = \omega_2$ given by Eq.\
(\ref{eq:symform}) is written in terms of the four independent 1-forms
of the unconstrained phase space $\Gamma_n = \Gamma_2$ of dimension $2n
= 4$.  When one imposes the Hamiltonian constraint $H=0$, one of the
four 1-forms appearing in $\omega_2$ can be written in terms of the
other three.  Thus one can write the symplectic structure 2-form in
terms of three 1-forms that are independent on the constraint
hypersurface $H=0$.  Choosing these three to be various combinations of
the differentials of $\alpha$, $\dot{\alpha}$, $\varphi$,
$\dot{\varphi}$, $\beta$, $\theta$, and $M$, one can write the 2-form on
the constraint hypersurface as
\begin{eqnarray}
\omega &=& S_0 e^{3\alpha} \left(
  -\frac{d\dot{\varphi}}{d\alpha} d\alpha \wedge d\varphi
  +\frac{d\varphi}{d\alpha} d\alpha \wedge d\dot{\varphi}
  - d\varphi \wedge d\dot{\varphi} \right)
       \nonumber \\
&=& S_0 e^{5\alpha} \left(
  -\ddot{\varphi} d\dot{\alpha} \wedge d\varphi
  +\dot{\varphi} d\dot{\alpha} \wedge d\dot{\varphi}
  -\ddot{\alpha} d\varphi \wedge d\dot{\varphi} \right)
               \nonumber \\
&=& S_0 e^{3\alpha-2\beta} \left(
   \frac{d\theta}{d\alpha} d\alpha \wedge d\beta
   -\frac{d\beta}{d\alpha} d\alpha \wedge d\theta
   + d\beta \wedge d\theta \right)
               \nonumber \\
&=& S_0 e^{5\alpha-2\beta} \left(
   \dot{\theta} d\dot{\alpha} \wedge d\beta
   -\dot{\beta} d\dot{\alpha} \wedge d\theta
   +\ddot{\alpha} d\beta \wedge d\theta \right)
               \nonumber \\
&=& \frac{1}{2} S_0 \left(
   -\frac{d\theta}{d\alpha} d\alpha \wedge dM
   +\frac{dM}{d\alpha} d\alpha \wedge d\theta
   -dM \wedge d\theta \right).
\label{eq:omega}
\end{eqnarray}
Here the derivatives with respect to time (with $n=1$) or to $\alpha$
that are the coefficients of the basis 2-forms inside the parentheses
are derivatives along the trajectories forming the cosmological
spacetime solutions, unlike the basis 1-forms that make up the basis
2-forms that are differentials transverse to the trajectories.

On an initial data surface that is an extremum of the scale size or of
the logarithm of the rescaled scale size, $\alpha = \alpha_m$, where
$\dot{\alpha} = 0$, one has $\beta = \beta_m$, $\theta = \theta_m$, and
$M = M_m = e^{3\alpha_m -2\beta_m}$, and from the two coordinates
$\beta_m$ and $\theta_m$ on this initial data surface that I shall call
$S_e$, one gets $\alpha_m = -(1/2)\ln{(\lambda - e^{-2\beta_m})}$ or
$\beta_m = -(1/2)\ln{(e^{-2\alpha_m} - \lambda)}$, then giving $M_m =
e^{\beta_m}(1+\lambda e^{2\beta_m})^{-3/2} = e^{\alpha_m}(1-\lambda
e^{2\alpha_m})$.  Note that for an extremum we must have $\alpha_m \leq
-(1/2)\ln{\lambda}$ or $a_m \leq 1/\sqrt{\lambda}$ or $\lambda a_m^2
\leq 1$, but $\beta_m$ can be an arbitrary real number (though one no
longer has the full range of all real numbers for $\beta$ if the
extremum is required to be a global minimum for $a$ or $\alpha$).  One
can alternatively label the initial data surface $S_e$ of extrema (all
points in the constrained hypersurface $H=0$ where also $\dot{\alpha} =
0$) by the two coordinates $a_m$ and $\theta_m$, both of which have
finite ranges, $0 < a_m \leq 1/\sqrt{\lambda}$ and $0 \leq \theta_m <
2\pi$.  Then on that initial data surface $S_e$ one has $a=a_m$, $\alpha
= \alpha_m = \ln{a_m}$, $\beta = \beta_m = -(1/2)\ln{(1/a_m^2 -
\lambda)}$, $\varphi = \varphi_m = e^{-\beta_m} \cos{\theta_m}$,
$\dot{\varphi} = \dot{\varphi}_m = e^{-\beta_m} \sin{\theta_m}$,
$\ddot{\alpha}_m = [1-3(1-\lambda a_m^2)\sin^2{\theta_m}]/a_m^2$, and
$M_m = a_m(1-\lambda a_m^2)$.

The pullback of $\omega_2$ to an initial data surface $S_e$ that is at
an extremum of $a$ and of $\alpha$, where $\dot{a} = 0$ and hence where
$a_m = a_m(\varphi_m,\dot{\varphi}_m) = (\lambda + \dot{\varphi}_m^2 +
\varphi_m^2)^{-1/2} = (\lambda + e^{-2\beta_m})^{-1/2} \leq
1/\sqrt{\lambda}$, is 
\begin{eqnarray}
\omega &=&
-S_0 e^{5\alpha_m} \ddot{\alpha}_m d\dot{\varphi}_m \wedge d\varphi_m
                      \nonumber \\
&=& S_0 e^{5\alpha_m-2\beta_m} \ddot{\alpha}_m d\beta_m \wedge d\theta_m
                      \nonumber \\
&=& S_0 \left[1-3\left(1-\lambda a_m^2\right)\sin^2{\theta_m}\right]
             da_m \wedge d\theta_m
                      \nonumber \\
&=& \mu_0 \left[1-3\left(1-x^2\right)\sin^2{\theta_m}\right]
             dx \wedge d\theta_m,
\label{eq:omega2}
\end{eqnarray}
where $\mu_0 \equiv S_0/\sqrt{\lambda} \approx 3\times 10^{67}$ and $x
\equiv \sqrt{\lambda}a_m$, which has the range $0 < x \leq 1$.

Because both $a_m$ (or $x$) and $\theta_m$ have finite ranges, and
because the integrand is bounded above within this range, the measure
for the set of solutions with a nonzero extremum for $a$ is finite. For
the case with zero cosmological constant, Hawking and I showed \cite{HP}
that the solutions with an extremum within a finite range of $a$ have
finite measure, but for that model there is no upper bound on $a$ at an
extremum, and almost all solutions have a maximum for $a$, so the total
measure for solutions with maxima is infinite.  But in the present case,
the positive cosmological constant imposes an upper limit on the value
of $a$ at an extremum.

Therefore, we see that the set of solutions with a nonzero extremum for
$a$ has a finite canonical measure, out of the infinite measure for all
solutions for a $k=+1$ FLRW cosmology with a minimally coupled massive
scalar field and a positive cosmological constant.  The finite measure
of solutions includes both totally nonsingular solutions, which have a
nonzero global minimum for $a$, and also solutions with both a big bang
and a big crunch, which have a finite global maximum for $a$.  It also
includes solutions that start at a big bang and eventually expand
forever, and their time reverses that contract from $a=\infty$ to a big
crunch, so long as they have a local extremum for $a$, where $\dot{a}$
or the Hubble variable $\dot{\alpha}$ is zero.  The only set of
solutions that have infinite measure are those that expand monotonically
from a big bang at $a=0$ to infinite size at $a=\infty$, or the time
reverses that contract monotonically from $a=\infty$ to a big crunch at
$a=0$.

\section{Canonical measure for nonsingular cosmologies}

We have seen that the total Liouville-Henneaux-Gibbons-Hawking-Stewart
canonical measure for nonsingular
Friedmann-Lema\^{\i}tre-Robertson-Walker cosmologies with a minimally
coupled massive scalar field is finite.  (For there to be a nonzero
measure for such nonsingular cosmological solutions of the
Einstein-scalar field equations, we need a closed cosmology with $k=+1$
to allow $a$ to have a minimum value, and we need a positive
cosmological constant to allow $a$ to go to infinity asymptotically in
both directions of time.)  Now let us calculate the measure for the
nonsingular solutions.

The canonical Liouville-Henneaux-Gibbons-Hawking-Stewart measure or
volume (area) element $\omega$ on the initial data surface $\Gamma_1$
with $\dot{a}=1$, given by Eq.\ (\ref{eq:omega2}), has a sign of the
coefficient of $dx \wedge d\theta_m$ that is proportional to $\ddot{a} =
\ddot{a}_m = [1-3(1-\lambda a_m^2)\sin^2{\theta_m}]/a_m =
\sqrt{\lambda}[1-3(1-x^2)\sin^2{\theta_m}]/x$, the acceleration of the
scale factor $a$ at its extremum.  When this is positive, the extremum
is a local minimum for the scale factor; when $\ddot{a}_m < 0$, the
extremum is a local maximum for $a$.  If we integrate $\omega$ over the
range giving positive $\ddot{a}_m$, we get the finite measure $\mu_1 =
(4\pi/\sqrt{27})\mu_0 \approx 2.4184\mu_0$.  If we reverse the sign of
$\omega$ and integrate it over the range giving negative $\ddot{a}_m$,
we get the same finite measure, $\mu_2 = (4\pi/\sqrt{27})\mu_0$. 
Therefore, the total measure for solutions with extrema for $a$ is not
greater than $\mu_3 = \mu_1 + \mu_2 = (8\pi/\sqrt{27})\mu_0$, finite.

However, solutions may have more than one extremum for $a$, and $\mu_3$
counts all such solutions with a multiplicity given by the number of
extrema that they have.  Therefore, let us calculate what the measure is
for nonsingular solutions by just taking the measure at the nonzero
global minimum for $a$, avoiding the overcounting of a multiplicity of
minima.  For this calculation we shall assume that $\lambda \ll 1$,
as it indeed is in our part of the universe where $\lambda \equiv
\Lambda/(3m^2) \approx 5\times 10^{-111}$, and hence drop correction
terms proportional to positive powers of $\lambda$.

Most of the measure for the nonsingular solutions will come from values
of $a$ not too much less than the maximum value for an extremum, which
is at $a=1/\sqrt{\lambda}$ or $x \equiv \sqrt{\lambda}a_m = 1$. 
Therefore, we can assume that $x$ is not enormously smaller than unity
for almost all of the measure.  For a nonsingular solution that has a
global minimum at $\alpha \equiv \ln{a} = \alpha_m = \ln{x} -
(1/2)\ln{\lambda} \gg 1$ (with $-(1/2)\ln{\lambda} \approx 127 \gg
1$), the fact that Eq.\ (\ref{eq:betadot}) implies that $\beta$ cannot
decrease as $\alpha$ increases implies that the Hamiltonian constraint
equation Eq.\ (\ref{eq:bc}) gives
\begin{equation}
\dot{\alpha}^2 = \lambda + e^{-2\beta} - e^{-2\alpha}
               \leq \lambda + e^{-2\beta_m} - e^{-2\alpha} 
               = e^{-2\alpha_m} - e^{-2\alpha}
               \leq \frac{\lambda}{x^2}.
\label{eq:bci}
\end{equation}
For $x \gg \sqrt{\lambda} \approx 7\times 10^{-56}$, we thus get
$\dot{\alpha}^2 \ll 1$.

As a result, Eq.\ (\ref{eq:thetadot}) implies that $\dot{\theta} \approx
-1$ to high accuracy, and Eq.\ (\ref{eq:Mdot}) implies that $M \equiv
e^{3\alpha-2\beta} \equiv a^3(\varphi^2 + \dot{\varphi}^2)$ stays very
nearly constant along most of the nonsingular trajectories.  Eq.\
(\ref{eq:betadot}) implies that $\dot{\beta}$ is also very small, so
over a number of oscillations of the scalar field that is not too large
(a change in the phase $\theta$ that is not too many times $2\pi$),
neither $\alpha$ nor $\beta$ change much, though after a very long time
and a huge number of oscillations of the scalar field (enormous change
in $\theta$), both $\alpha$ and $\beta$ grow indefinitely, while $M$ and
$\psi = \theta+t$ stay nearly constant and indeed both approach precise
constants in the infinite future, $M_\infty$ and $\psi_\infty$.  

(To define $\psi_\infty$ unambiguously, set $t=0$ at the global minimum
for $a$ and require $0 \leq \theta_m < 2\pi$ there.  One can make this
definition not only for nonsingular solutions but also for big bang
solutions that start at global minimum for $a$ that is $a=1$, where one
can set $t=0$, and then evolve to infinite $a$ where $\psi_\infty$ can
be evaluated.  To circumvent the jumps in $\theta_m$ at the minimum that
would occur with a sequence of solutions with $\theta_m$ approaching
$2\pi$ and then jumping back to 0, instead of defining the two real
constants $M_\infty$ and $\psi_\infty$, it would be better to define the
one complex constant $Z = \sqrt{M_\infty}e^{i\psi_\infty}$, which is
invariant under shifting $\theta_m$ and hence $\psi$ and $\psi_\infty$
by an integer multiple of $2\pi$.  Any solution that evolves to $a =
\infty$ will have a definite value for $Z$ that may be obtained by
analytic integration of the equations of motion from any initial point
in the constrained phase space, except for the nonsingular solutions
that have two equal global minima for $a$ and therefore the ambiguity of
which one to use for setting the zero of $t$, and the solutions that are
the limits of sequences of solutions with bounces of arbitrarily large
values of the scalar field \cite{dnpcqg}.  Both of these types of
particular solutions will have zero measure, so all but a set of measure
zero of the solutions that evolve to $a=\infty$ will be integrable,
having two real constants of motion that may be given by one complex
constant $Z$, that are analytic functions over all but a set of
hypersurfaces of measure zero in the constrained phase space.  The same
will be true for $k=0$ and $k=-1$ FLRW-scalar models with a nonnegative
cosmological constant to allow solutions to expand to $a=\infty$, though
since they cannot have extrema of $a$ that in the $k=+1$ case can lead
to hypersurfaces of the constrained phase space where the constants of
motion are not analytic, it appears that the $k=0$ and $k=-1$ FLRW-scalar
models will be totally integrable over the entire constrained phase
space.  In fact, since these models give $a$ expanding monotonically
from $a=0$ to $a=\infty$, or the time reverse, one can not only define
constants of motion by the asymptotic behavior of $M$ and $\psi$ at
$a=\infty$ that gives rise to the complex constant $Z$, but also by the
asymptotic behavior at $a=0$, such as the value of $v-u$ where either
one of these null coordinates $u$ and $v$ goes to zero, and the value of
the slope $dv/du$ there.)

During the oscillations of the scalar field while $\alpha$ and $\beta$
stay near their values at the extremum, one can write
\begin{equation}
\frac{d^2\alpha}{d\theta^2} \approx \frac{d^2\alpha}{dt^2}
\approx
\frac{\lambda}{2x^2}
\left[3x^2-1+3(1-x^2)(1-\cos{2\theta})\right].
\label{eq:ddalpha}
\end{equation}
Integrating this gives
\begin{equation}
\frac{d\alpha}{d\theta} \approx \frac{\lambda(3x^2-1)}{2x^2}
\left[\theta-\theta_0+B\sin{2\theta})\right],
\label{eq:dalpha}
\end{equation}
where
\begin{equation}
B \equiv \frac{3(1-x^2)}{2(3x^2-1)},\ 
\theta_0 \equiv \theta_m + B\sin{2\theta_m},
\label{eq:Btheta_0}
\end{equation}
and then finally
\begin{equation}
\alpha \approx \alpha_m + \frac{\lambda(3x^2-1)}{4x^2}
\left[(\theta-\theta_0)^2-B\cos{2\theta}
     -(\theta_0-\theta_m)^2+B\cos{2\theta_m}\right],
\label{eq:alpha}
\end{equation}

One can then see that for $\theta=\theta_m$ to be not only a local
minimum (which requires merely $3x^2-1+3(1-x^2)(1-\cos{2\theta_m}) \geq
0$) but also a global minimum, one needs that $3x^2-1 \geq 0$ for a
nonnegative coefficient of the quadratic term in $\theta$.  Furthermore,
by sketching the behavior of $\alpha(\theta)$, one can see that for
$-\pi < 2\theta_0 < \pi$, one needs $-\pi/2 < 2\theta_m < \pi/2$; for 
$\pi < 2\theta_0 < 3\pi$, one needs $3\pi/2 < 2\theta_m < 5\pi/2$; for
$3\pi < 2\theta_0 < 5\pi$, one needs $7\pi/2 < 2\theta_m < 9\pi/2$; etc.
Thus we cannot have a global minimum with $\pi/2 < 2\theta_m < 3\pi/2$, 
$5\pi/2 < 2\theta_m < 7\pi/2$, etc.

If we choose $\theta_m$ to lie in the range $0 \leq \theta_m < 2\pi$,
then there are four allowed ranges of $\theta_m$ of width $\pi/4$
(covering half the full circle for $\theta_m$; the other half does not
give extrema that are global minima) that give equal contributions to
the measure.  Let us focus on the first, which is that part of $0 \leq
\theta_m < \pi/4$ that gives $\theta_0 \equiv \theta_m +
B\sin{2\theta_m} < \pi/2$.  Using Eq.\ (\ref{eq:Btheta_0}) to express
$B$ in terms of $x \sqrt{\lambda} a_m$ allows one to convert this to a
restriction on $x$ for $0 \leq \theta_m < \pi/4$:
\begin{equation}
x_m(\theta_m) \equiv 
\sqrt{
\frac{\pi-2\theta_m+3\sin{2\theta_m}}{3\pi-6\theta_m+3\sin{2\theta_m}}}
\leq x \leq 1.
\label{eq:xm}
\end{equation}
   
If we now integrate the canonical
Liouville-Henneaux-Gibbons-Hawking-Stewart measure or volume (area)
element $\omega$ in Eq.\ (\ref{eq:omega2}) over the initial data
surface, say $S_m$, that has $a$ not only an extremum but also a global
minimum (which is 4 times the integral of $\omega$ over the one region
above, in order to include all possibilities for $0 \leq \theta_m <
2\pi$ which give a global minimum), we get (using $y=2\theta_m$)
\begin{equation}
\mu_m \equiv \gamma \mu_0 = \int_{S_m} \omega
= \mu_0\left\{2-\int_0^{\pi/2} dy 
\left[(3\cos{y}-1)x_m
      +(1-\cos{y})x_m^3\right]\right\}. 
\label{eq:nonsingmeas}
\end{equation}

Doing the integral numerically with Maple 12 gave $\gamma \approx
0.86334$.  Putting in the numbers given above for $S_0 \equiv
(3\pi)/(2Gm^2) = (3\pi/2)(m_{\mathrm{Pl}}/m)^2 \approx 2\times 10^{12}$,
$\lambda \equiv \Lambda/(3m^2) \equiv 1/(mb)^2 \approx 5\times
10^{-111}$, and $\mu_0 = S_0/\sqrt{\lambda} \approx 3\times 10^{67}$
gives the measure for the nonsingular $k=+1$ FLRW cosmologies with the
observed value of the cosmological constant and a scalar field mass of
$m \approx 1.5\times 10^{-6}m_{\mathrm{Pl}}$ as
\begin{equation}
\mu_m \equiv \gamma\mu_0 \equiv \gamma\frac{S_0}{\sqrt{\lambda}} \equiv
\gamma\frac{3\sqrt{3}\pi}{2Gm\sqrt{\Lambda}} 
\approx 0.86334\mu_0 \approx 3\times 10^{67}. 
\label{eq:nonsingmeas2}
\end{equation}
To convert this to a number of quanta, say $N_m$, one divides the phase
space measure by $h = 2\pi\hbar = 2\pi$ in our units with $\hbar = c =
1$ to get
\begin{equation}
N_m \equiv \frac{\mu_m}{2\pi} \equiv \frac{\gamma\mu_0}{2\pi}
    \equiv \gamma\frac{S_0}{2\pi\sqrt{\lambda}} 
    \equiv \gamma\frac{3\sqrt{3}}{4Gm\sqrt{\Lambda}}
    \approx 0.13740 \mu_0 
    \approx 4\times 10^{66}. 
\label{eq:nonsingN}
\end{equation}

One can compare this with the maximum number of scalar field quanta, say
$N_M$, that one can have for a nonsingular $k=+1$ FLRW cosmology with a
positive cosmological constant if the scalar field acted as pressureless
dust, which is how it does act at late times when one averages over an
integer number of oscillations of the scalar field.  Then $M$ would stay
constant.  For a universe with a minimum value of $a$ that is $a_m$, one
gets $M = a_m(1-\lambda a_m^2)$, which has a maximum value (when
$1-3\lambda a_m^2 \equiv 1-3x^2 = 0$) of $M_M = 2/\sqrt{27\lambda}$. 
The physical energy density $(1/2)[(m\phi)^2+(\dot{\phi}/N)^2]=[3/(8\pi
G)]m^2(\varphi^2+\dot{\varphi}^2)$ (with $N=n/m=1/m$) multiplied by the
physical 3-volume $2\pi^2 A^3 = 2\pi^2 a^3/m^3$ then gives a physical
`mass' $\mathcal{M} = (S_0/2) m M$, so the maximum number of scalar dust
quanta of mass $m$ in a nonsingular $k=+1$ FLRW cosmology with a
positive cosmological constant is
\begin{equation}
N_M = \frac{1}{2} S_0 M_M
    = \frac{S_0}{\sqrt{27\lambda}}
    = \frac{\mu_0}{\sqrt{27}}
    \approx 0.19245 \mu_0 \approx 1.400607 N_m \approx 6\times 10^{66}.
\label{eq:nonsingNM}
\end{equation}

Therefore, the number of quanta corresponding to the actual phase space
measure over the nonsingular $k=+1$ FLRW cosmologies is
$\sqrt{27}\gamma/(2\pi) \approx 0.713976$ times the maximum of that for
a dust model with the same particle mass.  In the actual scalar field
model, the scalar field undergoes coherent oscillations in which the
phase has gravitational consequences, so it cannot be accurately modeled
by assuming that the scalar field is in a precise number eigenstate with
a totally undetermined phase, which would give zero pressure for a
homogeneous field such as is being assumed here.

\section{Canonical measure for inflationary cosmologies}

Nearly all of the finite total
Liouville-Henneaux-Gibbons-Hawking-Stewart canonical measure for
nonsingular Friedmann-Lema\^{\i}tre-Robertson-Walker cosmologies with a
minimally coupled massive scalar field and a positive cosmological
constant occurs for solutions that are not large deviations from empty
de Sitter spacetime.  For example, if one defines the rationalized
dimensionless energy density to be
\begin{equation}
\rho \equiv e^{-2\beta} \equiv \varphi^2 + \dot{\varphi}^2
     \equiv \frac{M}{a^3},
\label{eq:rho}
\end{equation}
which is $(8\pi G)/(3m^2)$ times the physical energy density $\hat{\rho}
= (1/2)[(m\phi)^2+(\dot{\phi}/N)^2]=[3/(8\pi
G)]m^2(\varphi^2+\dot{\varphi}^2) = [3/(8\pi G)]m^2\rho$ of the scalar
field, and sets $\rho = \rho_m = a_m^{-2} - \lambda$ at the global
minimum for the scale factor at $a = a_m$, then under the approximation
that $M$ is constant, one has $1/\sqrt{3\lambda} \leq a_m \leq
1/\sqrt{\lambda}$ and $\rho \leq \rho_m \leq 2\lambda$ everywhere in the
spacetime, so the physical energy density of the scalar field is never
more than twice the physical energy density $\hat{\rho}_\Lambda =
\Lambda/(8\pi G) = [3/(8\pi G)]m^2\lambda$ corresponding to the
cosmological constant, that is, $\hat{\rho} \leq 2\hat{\rho}_\Lambda$.

If $\rho_m = a_m^{-2} - \lambda > 2\lambda \ll 1$ at an extremum of $a$,
then for $\rho_m \ll 1$, $\dot{\alpha}^2 \ll 1$ for the entire solution,
so $M$ stays nearly constant at a value less than its maximum value for
nonsingular dust solutions ($M = M_M = 2/\sqrt{27\lambda}$), and the
resulting solutions collapse to $a=0$ rather than expanding to
infinity.  To obtain solutions with $\hat{\rho}_m > 2\hat{\rho}_\Lambda$
that expand to infinity in both directions of time rather than
collapsing to zero size, one need to have a period of inflation in which
$M$ grows larger to become larger than $M_M = 2/\sqrt{27\lambda}$, as
one can see from the following argument:

The Hamiltonian constraint equation in Eq.\ (\ref{eq:g}) can be written
in terms of $a$ and $M$ as
\begin{equation}
\dot{a}^2 = f(a) \equiv \lambda a^2 - 1 + \frac{M}{a}.
\label{eq:Mconstraint}
\end{equation}

After the end of a possible period of inflation (which requires $\rho
\stackrel{>}{\sim} 1 \gg \lambda$), during which $M$ can grow
exponentially, $M$ will become nearly constant as the scalar field
starts oscillating with a period much less than the inverse of the
Hubble expansion rate.  Then the universe will expand forever if $f(a)$
stays positive for all larger $a$.  For constant $M$, the extremum of
$f(a)$ is at $a = [M/(2\lambda)]^{1/3} = (M/M_M)^{1/3}/\sqrt{3\lambda}$.
The value at the extremum is $f(a) = (6.75 M^2\lambda)^{1/3} - 1 =
(M/M_M)^{2/3} - 1$, so if inflation ends before $a$ reaches the
extremum, one needs $M > M_M$ in order for $\dot{a}$ to stay positive as
$a$ passes through the extremum of $f(a)$.  If inflation gives $M \leq
M_M$, it will end far before this value of $a$ is reached, so one needs
inflation to give $M > M_M = 2/\sqrt{27\lambda} \gg 1$ if one starts at
an extremum with $\rho_m > 2\lambda$ and hence with $a_m <
1/\sqrt{3\lambda}$, where there are no noninflationary solutions ($M$
nearly constant) that expand to infinity in both directions of time and
hence are nonsingular.

For a symmetric bounce ($\dot{\varphi}_m=0$ or $\theta_m = 0$) at
$\varphi_m = \varphi_b \gg 1$, I have calculated numerically
\cite{Symbounce} that the number of e-folds of inflation $N$ (not be be
confused with the previous use of $N$ for the lapse function) is
\begin{equation}
N(\varphi_b) \approx \frac{3}{2}\varphi_b^2
+ \frac{1}{3}\ln{\varphi_b} -1.0653 
-\frac{3\pi^2-14}{36\varphi_b^2} -\frac{0.4}{\varphi_b^4},
\label{eq:mm}
\end{equation}
and that the asymptotic value of $M$ is
\begin{equation}
M_\infty(\varphi_b) \approx 0.1815\varphi_b^{-3}e^{3N(\varphi_b)}
\approx \frac{0.08914 e^{4.5\varphi_b^2}}
{12\varphi_b^2+3\pi^2-14+24/\varphi_b^2}.
\label{eq:nn}
\end{equation}
To give $M > M_M$, this requires $N > N_M \approx 44.27$.

For $\ddot{\alpha} = \lambda + \rho_m(1-3\sin^2{\theta_m}) \approx
\rho_m(1-3\sin^2{\theta_m}) > 0$, we need $|\sin{\theta_m}| <
1/\sqrt{3}$.  If this is is true for $\rho_m \gg 1$ and $\theta_m$ is
not too close to the boundary, then the equations of motion in this
inflationary regime will generally cause $\dot{\varphi}$ and hence
$\theta$ to decay to near zero, and then one will get roughly $N \sim
(3/2)\rho_m$ e-folds of inflation, though the coefficient $(3/2)$ will
become some $\theta_m$-dependent number that is $(3/2)$ only at
$\sin{\theta_m}=0$.  However, for a crude estimate of the measure for
different amounts of inflation, let us ignore this effect.  Then Eq.\
(\ref{eq:omega2}) with $a_m = (\lambda + \rho_m)^{-1/2} \approx
\rho_m^{-1/2}$ gives the measure of varying numbers $N$ of e-folds of
inflation as
\begin{eqnarray}
\mu &=& \int\omega
                      \nonumber \\
&=& -\frac{1}{2}S_0\int(\lambda + \rho_m)^{-5/2}
\left[\lambda+\rho_m(1-3\sin^2{\theta_m})\right]d\rho_m\wedge d\varphi_m
                      \nonumber \\
&\approx& S_0(1-3\sin^2{\theta_m})d(\rho_m^{-1/2})\wedge d\varphi_m
                      \nonumber \\
&=& (\sqrt{6} - \cos^{-1}{3^{-3/2}})S_0 \int d(\rho_m^{-1/2})
                      \nonumber \\
&\sim& (3 - \sqrt{3/2}\cos^{-1}{3^{-3/2}})S_0 \int d(N^{-1/2}).
                      \nonumber \\
&\sim& S_0 \int d(N^{-1/2}).
\label{eq:inflationmeasure}
\end{eqnarray}
The second expression on the right hand side (after $\int\omega$) uses
the very good approximation that $\lambda \ll \rho_m$ for the
inflationary values of $\rho_m$ that are at least of the order of unity
in our units that set effectively set $m=1$.  The third evaluates the
integral over $\theta_m$.  The fourth uses the approximation $N \sim
(3/2)\rho_m$ that is actually only true for $\sin{\theta_m} = 0$, so the
fifth drops the uncertain numerical coefficient.

Thus we see that the measure for at least $N$ e-folds of inflation is
proportional to $1/\sqrt{N}$ for large $N$.  If we take the fraction of
the total measure for nonsingular solutions, which was $\gamma\mu_0 =
\gamma S_0/\sqrt{\lambda} \approx 3\times 10^{67}$, the fraction of the
total measure for at least $N$ e-folds of inflation is
\begin{equation}
F \sim \frac{\sqrt{\lambda}}{\sqrt{N}}
 = \frac{\sqrt{\Lambda/3}}{m\sqrt{N}}
 \sim \frac{10^{-55}}{\sqrt{N}}
 \sim \frac{10^{-56}}{\sqrt{N/N_M}}.
\label{eq:fracinflationmeasure}
\end{equation}

That is, the fraction of the measure for all nonsingular $k=+1$ FLRW
cosmologies that have inflation (requiring $N > N_M \approx 44.27$ for
the present toy model with just the observed value of the cosmological
constant and a massive scalar field with $m \approx 1.5\times 10^{-6}
m_{\mathrm{Pl}}$; it would be higher if the scalar field could decay
into radiation at the end of inflation) is about $10^{-56}$.  However,
the fraction goes down with the minimum number of e-folds required, $N$,
only by an inverse square root of $N$, and not as $e^{-3N}$, so there is
no conflict with not observing the universe to have such a minimal
amount of inflation that spatial curvature is observable.

\section{Conclusions}

Although the total canonical Liouville-Henneaux-Gibbons-Hawking-Stewart
measure is infinite for Friedmann-Lema\^{\i}tre-Robertson-Walker
classical universes with a minimally coupled massive scalar field and a
positive cosmological constant, it is finite for completely nonsingular
solutions (which have positive scale factor everywhere).  Nearly all of
the solutions have the energy density never more than twice the
effective energy density of the cosmological constant, but the tiny
fraction, $\sim 10^{-56}$, of the measure in which the energy density
ever exceeds this tiny amount has at least $\sim 44$ e-folds of
inflation and gives a measure that decreases only very slowly (as an
inverse square root) with the minimal number of e-folds required.

\newpage

\section{Acknowledgments}

I am grateful for the hospitality of Princeton University, where I was
motivated to work on this problem by a talk by Neil Turok on his latest
approach to the measure issue \cite{Tur}.  This research was
supported in part by the Natural Sciences and Engineering Research
Council of Canada.


\end{document}